\documentclass[11pt,a4paper]{article}
\usepackage{jheppub}

\usepackage{amsmath}
\usepackage{amssymb}
\usepackage{mathrsfs}
\usepackage{graphicx}
\usepackage{color}
\usepackage{pstricks}

\newcommand{\grad}{\nabla}

\begin{document}

\title{Rotating Black Droplet}

\author[a]{Sebastian Fischetti}
\author[a]{Jorge E. Santos}

\affiliation[a]{Department of Physics \\ University of California,
Santa Barbara \\ Santa Barbara, CA 93106, USA }

\emailAdd{sfischet@physics.ucsb.edu}
\emailAdd{jss55@physics.ucsb.edu}


\arxivnumber{1304.1156}

\abstract{
We construct the gravitational dual, in the Unruh state, of the ``jammed'' phase of a CFT at strong coupling and infinite $N$ on a fixed five-dimensional rotating Myers-Perry black hole with equal angular momenta. When the angular momenta are all zero, the solution corresponds to the five-dimensional generalization of the solution first studied in \cite{Figueras:2011va}. In the extremal limit, when the angular momenta of the Myers-Perry black hole are maximum, the Unruh, Boulware and Hartle-Hawking states degenerate. We give a detailed analysis of the corresponding holographic stress energy tensor for all values of the angular momenta, finding it to be regular at the horizon in all cases.  We compare our results with existent literature on thermal states of free field theories on black hole backgrounds.
}

\maketitle

\section{Introduction}
\label{sec:intro}

Since the revolutionary discovery that black holes radiate with an almost thermal spectrum, much effort has been devoted to gaining a deeper understanding of this process.  While considerable advances have been made since Hawking's seminal paper~\cite{Hawking1974}, there are still ample open issues that remain to be addressed.  In particular, most of the work regarding Hawking radiation has focused on free fields; studying the behavior of \textit{interacting} fields around a black hole is more challenging.  Though some progress has been made with weakly interacting fields using perturbative methods~\cite{BirrellDavies}, strongly interacting fields pose a much more difficult problem.

Fortunately, the AdS/CFT correspondence~\cite{Maldacena:1997re} provides an invaluable tool for tackling this problem.  Early applications of holography to studying Hawking radiation coupled the field theory to dynamical gravity, which is holographically dual to Randall-Sundrum braneworld models~\cite{Randall:1999vf,Tanaka:2002rb,Emparan:2002px}.      However, if one is not interested in the backreaction of the Hawking radiation on the spacetime, the CFT can be placed on a nondynamical background spacetime~$\partial\mathcal{M}$.  Then AdS/CFT claims that the CFT is dual to a spacetime~$\mathcal{M}$ with conformal boundary~$\partial\mathcal{M}$, with~$\mathcal{M}$ a solution to classical Einstein gravity with negative cosmological constant.  While most applications of the duality place the field theory on ordinary Minkowski space, one may in principle select any conformal boundary structure one desires.  In particular, in order to study Hawking radiation in the CFT,~\cite{Hubeny:2009ru,Hubeny:2009rc} considered a CFT living on fixed~$d$-dimensional background containing a black hole of size~$R$ and temperature~$T_\mathrm{BH}$.  Far away from the black hole, the CFT is in a thermal state with a prescribed temperature~$T_\infty$.  From the perspective of the CFT, the black hole and the plasma at infinity act as heat sources and sinks, and one can study the exchange of heat between them.

In fact, a complete picture requires the introduction of another parameter~$T_0$, representing the temperature of the CFT plasma near the black hole.  Though it might seem natural to take this temperature to be equal to that of the black hole, it is possible to set~$T_0 \neq T_\mathrm{BH}$ at the expense of making the CFT stress tensor singular at the black hole horizon.  Details of this so-called ``detuning'' can be found in~\cite{Fischetti:2012vt}, though in our solution we will only consider the ``tuned'' case~$T_0 = T_\mathrm{BH}$ and will therefore never introduce the parameter~$T_0$ explicitly.

\begin{figure}[t]
\begin{center}
\hfill
\begin{pspicture}(-1,0)(16,5)
\psset{unit=0.95cm}


\pscustom{
\gsave
	\pscurve(0,2)(0.8,2.05)(1.2,2.2)(1.6,2.6)(2,4)
	\psline(5,4)
	\pscurve(5.4,2.6)(5.8,2.2)(6.2,2.05)(7,2)
	\psline(7,1.2)(0,1.2)(0,2)
	\fill[fillstyle=solid,fillcolor=lightgray]
\grestore
}

\psline(0,4)(7,4)
\psdots(2,4)(5,4)
\pscurve(0,2)(0.8,2.05)(1.2,2.2)(1.6,2.6)(2,4)
\pscurve(5,4)(5.4,2.6)(5.8,2.2)(6.2,2.05)(7,2)

\psline{|-|}(2,4.2)(5,4.2)
\rput[b](3.5,4.3){$2R$}
\psline{|-|}(-0.2,2)(-0.2,4)
\rput[r](-0.3,3){$T_\infty^{-1}$}

\psline{->}(6,3.4)(5.2,3.7)
\rput[l](6.2,3.4){Throat}
\psline{->}(2.6,1.6)(1.4,2.1)
\psline{->}(4.4,1.6)(5.6,2.1)
\rput[t](3.5,1.6){Shoulders}
\rput[b](0.9,4.05){Boundary}
\rput(3.5,3){Black Funnel}
\rput(3.5,0.5){$(a)$}


\pscustom{
\gsave
	\pscurve(12,4)(12.1,3.8)(12.3,3.6)(12.5,3.55)(12.7,3.6)(12.9,3.8)(13,4)
	\psline(12,4)
	\fill[fillstyle=solid,fillcolor=lightgray]
\grestore
}

\pscustom{
\gsave
	\pscurve(9,2)(10,2)(11,2)(12.5,2.2)(14,2)(15,2)(16,2)
	\psline(16,1.2)(9,1.2)
	\fill[fillstyle=solid,fillcolor=lightgray]
\grestore
}

\psline(9,4)(16,4)
\psdots(12,4)(13,4)
\pscurve(12,4)(12.1,3.8)(12.3,3.6)(12.5,3.55)(12.7,3.6)(12.9,3.8)(13,4)
\pscurve(9,2)(10,2)(11,2)(12.5,2.2)(14,2)(15,2)(16,2)

\psline{|-|}(12,4.2)(13,4.2)
\rput[b](12.5,4.3){$2R$}
\psline{|-|}(8.8,2)(8.8,4)
\rput[r](8.7,3){$T_\infty^{-1}$}

\psline{->}(13.4,3.4)(13,3.7)
\rput[t](14.2,3.3){Black Droplet}
\rput[b](9.9,4.05){Boundary}
\rput[t](12.5,1.8){(Deformed) Planar BH}
\rput(12.5,0.5){$(b)$}

\end{pspicture}
\caption{A sketch of the relevant solutions: {\bf (a):} black funnel and  {\bf (b):} black droplet above a deformed planar black hole. Both describe possible states of dual field theories in contact with heat baths at temperature $T_\infty$ on spacetimes containing black holes of horizon size $R$. The top line corresponds to the boundary, with the dots denoting the horizon of the boundary black hole.  The shaded regions are those inside the bulk horizons.}
\label{fig:dropfun}
\end{center}
\end{figure}
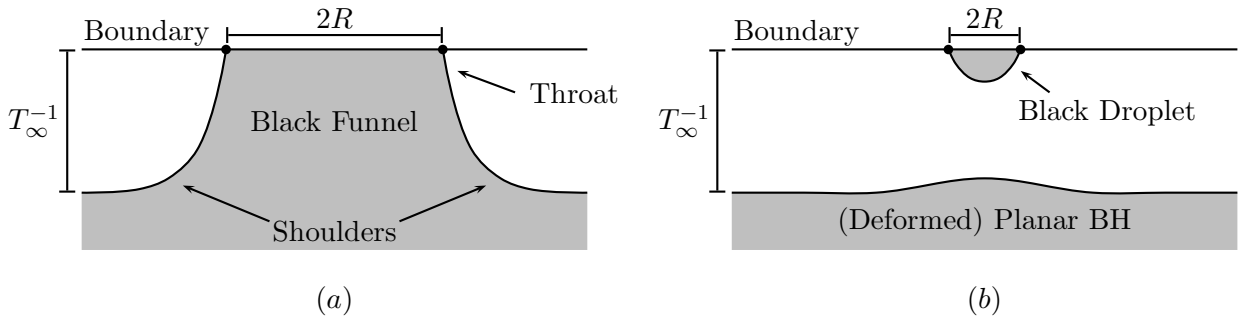

Regarding the dual geometry,~\cite{Hubeny:2009ru,Hubeny:2009rc} conjecture two possible families of solutions: so-called ``black droplets'' and ``black funnels,'' illustrated in Figure~\ref{fig:dropfun}.  In the black funnel solutions, the horizon of the boundary black hole is connected to an asymptotically planar horizon in the bulk; this connectedness manifests itself in the CFT as strong coupling between the boundary black hole and the heat bath at infinity, leading to an exchange of heat between them.  In the black droplet solutions, on the other hand, the horizon of the boundary black hole is disconnected from the asymptotically planar horizon in the bulk, implying that the boundary black hole is not coupled to the plasma at infinity (or rather, that the coupling is suppressed by~$\mathcal{O}(1/N^2)$ is this large-$N$ picture).  From the perspective of the CFT, the transition from funnels to droplets is reminiscent of many soft condensed matter systems which exhibit a transition from a fluid-like behavior to rigid behavior with no flow (e.g. sand in an hourglass, cars on a highway).  To borrow from the soft condensed matter nomenclature, we will refer to the the CFT transition from funnels to droplets as a ``jamming'' phase transition, and will denote the CFT dual of the droplet as a ``jammed'' phase\footnote{We thank Jean Carlson for introducing us to this terminology.}.

The authors of~\cite{Hubeny:2009ru,Hubeny:2009rc} postulate that this jamming transition might occur as the size~$R$ of the boundary black hole (or alternatively, as the temperature~$T_\infty$ of the heat bath at infinity) is varied.  Thus the dimensionless parameter~$RT_\infty$ should characterize which phase is thermodynamically preferred.

An analytic construction of black droplets and black funnels has only been performed in select few cases.  In~\cite{Hubeny:2009kz,Caldarelli:2011wa}, black droplets and funnels were constructed from the AdS C-metric, while~\cite{Fischetti:2012ps} constructed analytic funnels in~$d=2$ dual to the Unruh state of the CFT (black droplets cannot exist in~$d=2$).  However, many of the most interesting cases (notably, asymptotically flat boundary black holes in~$d > 2$) do not lend themselves to analytic construction, and one must resort to numerics.  To that end,~\cite{Santos:2012he} numerically constructed black funnels dual to the Hartle-Hawking state of the CFT, while~\cite{Fischetti:2012vt} constructed so-called global funnels to study the exchange of heat between two boundary black holes of different temperatures.

Slightly less effort has been made to construct black droplets, however.  Besides those obtained from the AdS C-metric, the only construction to date has been that of~\cite{Figueras:2011va}, which constructed a droplet dual to the Unruh state of a CFT on a~$d=4$ Schwarzschild background (hereafter generally denoted Schw$_d$).  In our notation, this solution sets~$T_\infty = 0$, so that the bulk geometry is that of a black droplet suspended above an extremal Poincar\'e horizon deep in the bulk.  The boundary stress tensor contains no flux terms, and the nonzero components of the stress tensor exhibit a~$1/r^5$ power-law falloff at large distances from the black hole.  This relatively rapid falloff is characteristic of the jammed phase, as it indicates that the CFT plasma is relatively well-localized near the black hole.  Interestingly, the stress tensor is regular on both the past and future horizons; the authors speculate that the inclusion of one-loop graviton corrections in the bulk would render the stress tensor singular on the past horizon, as is typical of Unruh states.

Perhaps some clarifying remarks are in order regarding our nomenclature of CFT states.  In free field theory, the distinction between the Unruh, Hartle-Hawking, and Boulware vacua is conventionally based on the behavior of the field theory stress tensor at the horizon and at null infinity.  In the Hartle-Hawking state, the stress tensor is regular on both the future and past horizon; in the Unruh state, the stress tensor is empty at past null infinity and regular on the future horizon; and in the Boulware state, the stress tensor is empty at both future and past null infinity.  However, one can also understand these behaviors very physically from the point of view of the field theory.  The Hartle-Hawking state is one in which the field theory is in thermal equilibrium, with all temperatures equal; in our language,~$T_\mathrm{BH} = T_0 = T_\infty$.  The Unruh state is one in which the temperature of the heat bath at infinity is taken to zero (so that for any nonzero~$T_0$, the black hole acts as a source for heat to flow to infinity);~$T_\infty = 0$.  Finally, the Boulware state is the state of minimum energy, requiring the temperature of the field theory to be zero everywhere;~$T_0 = T_\infty = 0$.  A crucial point to note is the appearance of the field theory temperature~$T_0$ (rather than the black hole temperature~$T_\mathrm{BH}$) in the above definitions.  Thus the regularity properties of the field theory stress tensor in a given state can be thought of as a consequence of how the temperatures~$T_0$ and~$T_\mathrm{BH}$ are tuned, and not as a defining feature of the state itself.  For instance, the ubiquitous singularity of the Boulware vacuum on the horizon of a nonextremal black hole is due to the detuning~$T_0 = 0 \neq T_\mathrm{BH}$, whereas the Hartle-Hawking state is generally regular on the black hole horizon because of the requirement that~$T_0 = T_\mathrm{BH}$.  When labeling states in strongly coupled CFTs, we will therefore operate under these latter definitions in terms of the temperatures~$T_0$ and~$T_\infty$ of the CFT.  For example, we would claim that the CFT dual to the droplet constructed in~\cite{Figueras:2011va} is in the Unruh state, but not the Boulware state, because~$T_\infty = 0$ but~$T_0 = T_\mathrm{BH} \neq 0$.  These definitions are also independent of whether the CFT is in the jammed or unjammed phase, so that the categorization of the state and phase of the CFT provide complementary descriptions of its behavior.

Our purpose in this paper is to generalize the result of~\cite{Figueras:2011va} by adding a new ingredient: rotation.  Though~\cite{Steif:1993zv,Hubeny:2009rc} compute the stress tensor on a rotating BTZ black hole background, the effect of rotation on a stress tensor in an \textit{asympototically flat} black hole spacetime has not been studied extensively.  One might naturally expect that the inclusion of rotation can be accomplished by generalizing the result of~\cite{Figueras:2011va} to a Kerr spacetime, but because the Kerr metric is cohomogeneity two, this would lead to a cohomogeneity three problem in the bulk.  Instead, we take the boundary spacetime to be the~$d=5$ equal-angular-momentum Myers-Perry metric~\cite{Myers1986304} (see also~\cite{Myers:2011yc} for a review), which is known to be cohomogeneity one and thus leads to a cohomogeneity two problem in the bulk.

The generalization to a non-extremal rotating black droplet should not be expected to exhibit very different behavior from its non-rotating counterpart: the boundary black hole still acts as a heat source at finite and nonzero temperature, and one studies the thermal coupling between the black hole and a zero-temperature heat bath at infinity.  However, the presence of rotation provides us with a new parameter to tune, which we can use to take the black hole to extremality; this will allow us to take~$T_0$ to zero without the need to perform any detuning.  For this reason, from this point on we will only refer to~$T_\mathrm{BH}$, with~$T_0 = T_\mathrm{BH}$ understood implicitly.

Taking~$T_\mathrm{BH}$ to zero with~$T_\infty$ fixed at zero can be thought of as some limit of an Unruh state.  Similarly, the fact that both the extremal black hole and the heat bath at infinity are at zero temperature places the CFT in the Boulware vacuum.  Finally, since in the extremal case~$T_\mathrm{BH} = T_\infty$, the CFT can be thought of as being in a Hartle-Hawking state (though perhaps it would be more correct to refer to this as a zero-temperature \textit{limit} of a Hartle-Hawking state).  Thus the extremal droplet is in some sense dual to a degenerate state which is a Boulware state and a limit of the Hartle-Hawking and Unruh states.

This paper is organized as follows.  In Section~\ref{sec:analytic}, we review the behavior of stress tensors on black hole spacetimes, and calculate the form our stress tensor must take on a Schw$_5$ background (i.e. when the angular momentum of the Myers-Perry black hole is taken to zero).  In Section~\ref{sec:numerics}, we briefly review the numerical method used and discuss the numerical construction of our solution.  In Section~\ref{sec:disc}, we compare our numerically extracted stress tensor to the expectations in Section~\ref{sec:analytic}, and discuss future directions to pursue.

{\bf Note}: In the final stages of this work we learned of \cite{FT}, which also constructs rotating black droplets and may have some overlap with our work.  Their paper will appear simultaneously with ours on the arXiv.

\section{A Review of Stress Tensors in Curved Spacetime}
\label{sec:analytic}

A common difficulty of field theories in curved spacetime is the lack of an unambiguous notion of ``particle.''  This poses no problem, however, if one instead limits oneself only to \textit{currents}.  As discussed in~\cite{Fulling1976}, when a physical system is probed, the objects that couple to probe fields are currents. In other words, the currents are the objects that appear in the interaction terms in the equations of motion of an interacting field.  In addition, a current is a genuinely local object and should retain a well-defined meaning even in the presence of strong external fields (such as spacetime curvature).  Thus most of the work devoted to understanding the process of radiation from black holes has focused not on quantum fields themselves, but on the renormalized vacuum expectation value (v.e.v.) of their stress-energy tensor~$\langle T^{\mu\nu} \rangle$.

In particular, the stress tensor of quantum fields can exhibit properties that are forbidden of classical stress tensors.  For instance, though a classical energy density must be non-negative, it is well-known that even in flat spacetime, the local energy density of a quantum field need not obey this same restriction\footnote{Here, we define the local energy density as measured by an observer with velocity~$u^\mu$ as
\[
\rho = u^\mu u^\nu \langle T_{\mu\nu} \rangle
\]}.  Perhaps the most famous example of a negative energy density is the Casimir effect~\cite{Casimir1948}, though the existence of negative energy densities in local quantum field theory is in fact quite general~\cite{Epstein1965}.

In the context of general relativity,~\cite{Fulling1976} studied the behavior of the energy-momentum tensor of a massless scalar field in a two-dimensional spacetime containing an accelerating mirror.  While the total energy radiated by the mirror was positive, a negative energy density and energy flux were present in the region near the mirror, whereas far from the mirror the stress tensor took the classical form of an outward flux of positive energy.  A similar result was found by~\cite{Davies1976}, which calculated the renormalized v.e.v. of the stress tensor of a massless scalar field in a two-dimensional model of black hole collapse.  As in the case of the radiating mirror, the near-horizon region of the newly-formed black hole was characterized by a negative energy density and a flux of negative energy flowing into the black hole, while far from the black hole the stress tensor again took the form of an outward flux of positive energy.  These results are consistent with Hawking's picture of particle-antiparticle creation near the horizon of a black hole, where negative-energy particles flow into the black hole while their positive-energy partners radiate to infinity: by conservation of energy, a positive-energy flux at infinity should be accompanied by a negative-energy flux through the black hole horizon.

These exact results were obtained by exploiting the conformal flatness of two-dimensional spacetimes.  Nevertheless, good approximations to thermal stress tensors have been obtained in higher dimensions.  Notably,~\cite{Page1982} used the Bekenstein-Parker Gaussian path integral approximation~\cite{Bekenstein1981} to find the approximate form of the stress tensor of a conformally invariant scalar field in the Hartle-Hawking state on the (four-dimensional) Schwarzschild background.  The behavior of the stress tensor was similar to the two-dimensional case: at infinity, the stress tensor behaved like a classical thermal stress tensor, while sufficiently near the horizon (for~$r \lesssim 2.34 M$) the energy density became negative.  These results were consistent with earlier numerical calculations~\cite{Candelas1980} of the stress tensor on the bifurcation two-sphere of Schwarzschild.  However, while one naturally expects a negative energy flux near the horizon in an Unruh state, it is not clear whether the ubiquity of negative energy densities near the horizon should extend to Hartle-Hawking states as well.

All of these results focused explicitly on conformally invariant noninteracting massless scalar fields.  One might therefore imagine that they have little bearing on the stress tensor we will obtain for our strongly coupled CFT.  In fact, the jammed phase dual to the droplet we construct cannot exist in a free field theory, so \textit{a priori} none of the results listed above can even approximately predict the behavior of the stress tensor we will extract.  Nevertheless, it is possible (and indeed, quite probable) that the stress tensor of quantum fields on black hole backgrounds should exhibit some kind of universal qualitative behavior.  This universal structure was studied in detail by Christensen and Fulling~\cite{Christensen1977}, who sought the general behavior of a static, spherically symmetric stress tensor on Schw$_4$ by solving the conservation equations
\begin{equation}
\label{eq:gradT}
\grad_\mu T^{\mu\nu} = 0
\end{equation}
directly, without restriction to any particular kind of field theory.  Their results are particularly useful to our case, because they are also valid in the case of a strongly coupled CFT.  In fact,~Schw$_5$ is just the zero-angular momentum limit of the Myers-Perry black holes that we Schwarzschild in higher dimensions.

\subsection{The General Static Spherically Symmetric Stress Tensor on Schw$_d$}
\label{subsec:SchwTmunu}

Consider the~$d$-dimensional Schwarzschild metric Schw$_d$ (also called the Tangherlini metric, after its discoverer~\cite{Tangherlini:1963bw}):
\begin{equation}
\label{eq:Schwd}
ds^2_\mathrm{Schw} = -f(r) \, dt^2 + \frac{dr^2}{f(r)} + r^2 \, d\Omega_{d-2}^2, \ \ \ f(r) \equiv 1-\left(\frac{r_0}{r}\right)^{d-3}.
\end{equation}
This spacetime is qualitatively identical to the familiar four-dimensional Schwarzschild, with horizon located at~$r = r_0$ and a temperature~$T = f'(r_0)/4\pi = (d-3)/4\pi r_0$.  Now, a static, spherically symmetric stress tensor must take the form
\begin{equation}
{T^\mu}_\nu = \begin{pmatrix} {T^t}_t & {T^t}_r & 0 \\ {T^r}_t & {T^r}_r & 0 \\ 0 & 0 & {T^\Omega}_\Omega \delta^i_j \end{pmatrix},
\end{equation}
where all components are functions of~$r$ only and the indices~$i$,~$j$ run over the angular coordinates.  Inserting this form into the conservation equations~\eqref{eq:gradT} and using the metric~\eqref{eq:Schwd}, we obtain the following differential equations:
\begin{subequations}
\begin{align}
0 &= \partial_r {T^r}_t + \frac{d-2}{r} \, {T^r}_t, \label{eq:dTrt} \\
0 &= \partial_r {T^r}_r + \left(\frac{d-2}{r} - \frac{d-3}{2r_0} \frac{(r_0/r)^{d-2}}{f(r)}\right){T^r}_r - \frac{d-3}{2r_0} \frac{(r_0/r)^{d-2}}{f(r)} \, {T^t}_t - \frac{d-2}{r} \, {T^\Omega}_\Omega. \label{eq:dTrr}
\end{align}
\end{subequations}
Equation~\eqref{eq:dTrt} can be immediately integrated to yield
\begin{equation}
\label{eq:Trt}
{T^r}_t = K\left(\frac{r_0}{r}\right)^{d-2}.
\end{equation}
To integrate equation~\eqref{eq:dTrr}, we first substitute~${T^t}_t = {T^\mu}_\mu - {T^r}_r - (d-2){T^\Omega}_\Omega$, and obtain
\begin{multline}
\label{eq:Trr}
{T^r}_r = \frac{(r_0/r)^{d-2}}{f(r)} \left[Q-K + \right. \\ \left. \frac{1}{2} \int_{r_0}^r \left( (d-3){T^\mu}_\mu(r') + (d-2) \left(2 (r'/r_0)^{d-3} - d + 1\right) {T^\Omega}_\Omega(r') \right) \frac{dr'}{r_0}\right].
\end{multline}
Equations~\eqref{eq:Trt} and~\eqref{eq:Trr} express the most general static, spherically symmetric stress tensor on Schw$_d$ in terms of two arbitrary constants~$Q$ and~$K$ and two arbitrary functions~${T^\mu}_\mu$ and~${T^\Omega}_\Omega$.  In order to study the physical behavior of these solutions, it will prove useful to define
\begin{subequations}
\begin{align}
\Theta(r) &\equiv {T^\Omega}_\Omega(r) - \frac{1}{2(d-2)} \, {T^\mu}_\mu(r), \\
G(r) &\equiv \frac{d-2}{2} \int_{r_0}^r \left(2(r'/r_0)^{d-3} - d + 1\right) \Theta(r') \, \frac{dr'}{r_0}, \label{eq:G} \\
H(r) &\equiv \frac{1}{2} \int_{r_0}^r \left(\frac{d-5}{2} + (r'/r_0)^{d-3}\right) {T^\mu}_\mu(r') \, \frac{dr'}{r_0}.
\end{align}
\end{subequations}
Converting to the tortoise coordinate~$dr_* = dr/f(r)$, the stress tensor can then be expressed as the sum of four pieces:
\begin{equation}
\label{eq:Tmunugeneral}
{T^\mu}_\nu = {(T_{(1)})^\mu}_\nu + {(T_{(2)})^\mu}_\nu + {(T_{(3)})^\mu}_\nu+ {(T_{(4)})^\mu}_\nu,
\end{equation}
where (in~$t$,~$r_*$ coordinates)
\begin{subequations}
\begin{align}
{(T_{(1)})^\mu}_\nu &= \mathrm{diag}\left\{-\frac{(r_0/r)^{d-2}}{f(r)} \, H(r) + \frac{1}{2} \, {T^\mu}_\mu(r), \, \frac{(r_0/r)^{d-2}}{f(r)} \, H(r), \, \frac{1}{2(d-2)} \, {T^\mu}_\mu(r) \, \delta^i_j \right\}, \\
{(T_{(2)})^\mu}_\nu &= K \, \frac{(r_0/r)^{d-2}}{f(r)} \begin{pmatrix} 1 & 1 & 0 \\ -1 & -1 & 0 \\ 0 & 0 & 0 \end{pmatrix},  \\
{(T_{(3)})^\mu}_\nu &= \mathrm{diag}\left\{-\frac{(r_0/r)^{d-2}}{f(r)} \, G(r) - (d-2)\Theta(r), \, \frac{(r_0/r)^{d-2}}{f(r)} \, G(r), \, \Theta(r) \, \delta^i_j \right\}, \label{eq:T3} \\
{(T_{(4)})^\mu}_\nu &= Q \, \frac{(r_0/r)^{d-2}}{f(r)} \, \mathrm{diag}\left\{-1, \, 1, \, 0 \right\}.
\end{align}
\end{subequations}
By converting to ingoing Eddington-Finkelstein coordinates~$dv = dt + dr_*$, it is straightforward to show that the stress tensor is regular on the future horizon only if~$Q$ vanishes\footnote{Regularity on the past horizon requires~$Q = 2K$.}.  Thus the above decomposition is convenient because each of the~$T_{(i)}$ isolates some interesting physical behavior of the stress tensor:~$T_{(1)}$ is the only of the~$T_{(i)}$ with a nonzero trace (and in fact, it is only a function of the trace~${T^\mu}_\mu$);~$T_{(2)}$ contains nonzero flux terms~${T^{r_*}}_t$;~$T_{(3)}$ is the only of the~$T_{(i)}$ which is both traceless and has nonzero tangential pressure components; and~$T_{(4)}$ is singular on the future horizon.

Christensen and Fulling proceed to make use of these results to study the behavior of the stress tensor of fields in the Unruh, Hartle-Hawking, and Boulware vacua on Schw$_4$.  Our goal, however, is less broad: we only wish to gain some insight on the v.e.v.~of the stress tensor of the jammed phase of the CFT.  To that end, we make the following observations regarding our droplet:
\begin{itemize}
\item The trace anomaly of any odd-dimensional CFT vanishes.  Since we work in~$d = 5$, we expect~$T_{(1)}$ to make no contribution to our stress tensor.
\item According to the arguments reviewed briefly at the end of Section~\ref{sec:intro}, we expect black droplets to be dual to a jammed phase of the CFT.  Such phases are characterized by a suppressed exchange of heat between the black hole and the thermal plasma at infinity, so we should expect our solution to radiate no flux.  From~\eqref{eq:Trt}, it is clear that the total flux radiated to infinity from the black hole is a constant proportional to~$K$, so we conclude that~$K = 0$ and so~$T_{(2)}$ makes no contribution to our stress tensor.
\item Unlike the past horizon, which is present only in maximal analytic extensions of black hole spacetimes, the future horizon is a genuine physical location present in any realistic model of gravitational collapse.  We thus expect physical quantities to be regular there.  In particular, we require the stress tensor of the jammed CFT to be regular there; this implies that~$Q = 0$, so that~$T_{(4)}$ does not contribute to the stress tensor either (incidentally, since we also have~$K = 0$, this means that the stress tensor will be regular at the past horizon as well).
\end{itemize}
We therefore conclude that only contribution to the stress tensor of the CFT state dual to the Schw$_5$ droplet should come from~\eqref{eq:T3}, and so should only depend on the one function~$\Theta(r)$.

In fact, we can go further and make claims about the behavior of~$\Theta(r)$.  The free field results summarized earlier in this section should not all apply to our stress tensor, but we might draw certain universal behavior from them to make a guess at the behavior of~$\Theta(r)$ (which we will then verify with our numerics).  First, note that the negative energy density near the horizon present in free field theories is an indication of the highly non-classical nature of the field there; only far from the black hole does the stress tensor of the free fields become classical.  But the jammed phase of the CFT is a highly non-classical state, as it consists of a plasma ``halo'' surrounding the black hole.  In analogy with the free field theory results, we might therefore conjecture that the energy density of the strongly coupled CFT becomes negative near the horizon as well.  In that case, we expand~${(T_{(3)})^t}_t = -3\Theta(r_0)/2 + \mathcal{O}(r-r_0)$ to see that a negative energy density near the horizon implies that~$\Theta$ is negative there.

Far from the black hole, one can argue that due to the weak coupling between the black hole and the heat bath at infinity, the components of the stress tensor should fall off ``rapidly''.  This statement can be quantified by following the logic of~\cite{Figueras:2011va}: there, the authors find that the v.e.v. of a CFT dual to a~$d = 4$ Schwarzschild droplet exhibits a~$1/r^5$ falloff.  In order to explain this falloff, the authors invoke the results of~\cite{Giddings:2000mu}, in which the linearized gravitational field created by a point-like source is calculated in the context of the Randall-Sundrum single braneworld model.  By using the behavior of the gravitational field far from the source, Einstein's equations
\begin{equation}
R_{\mu\nu} - \frac{1}{2}\, R g_{\mu\nu} = 8\pi G_N \langle T_{\mu\nu} \rangle
\end{equation}
can be used to obtain the expected large-$r$ behavior of the stress tensor.  For the case of a~$d = 4$ dimensional spacetime, the predicted falloff of the components~$\left\langle {T^\mu}_\nu \right\rangle$ goes like~$1/r^5$, consistent with the results of~\cite{Figueras:2011va}.  One can generalize this argument to show that in~$d=5$ dimensions, we expect a~$1/r^7$ falloff\footnote{We greatly thank the authors of~\cite{FT} for finding a mistake in an earlier version of this paper, where we incorrectly claimed a falloff of~$1/r^6$ in five dimensions.}.  This implies that for our solution, we should have~$\Theta(r) \sim 1/r^7$ and~$G(r) \sim 1/r^4$ at large~$r$.

\subsection{A Little Spin}
\label{subsec:spin}

Finally, let us consider the effect of giving the boundary black hole a nonzero angular momentum, thus ``spinning'' the droplet.  Unfortunately, an analysis similar to that performed for the nonspinning case is not illuminating, because even with the conservation equations~\eqref{eq:gradT}, there are too many independent functions in the stress tensor for a general solution to be tractable.  Nevertheless, we can make some qualitative claims based on physical arguments.

First, we clearly expect there to be angular flux terms as the CFT plasma is dragged around the spinning black hole.  This flux density should be maximal near the horizon, and decrease monotonically away from the black hole.  Similarly, the fact that the CFT plasma is being forced to rotate around the black hole leads us to expect that a centrifugal barrier forms around the black hole.  Indeed, by considering timelike geodesics in the equatorial plane of the five-dimensional equal-angular-momentum Myers-Perry black hole, we can check that the centrifugal barrier of the effective radial potential grows and moves away from the horizon as the spin of the black hole is increased.  This barrier may act to confine the plasma near the black hole horizon, effectively acting as a box around the black hole.  We might expect that this effect would present itself as an increase in the radial and tangential pressures near the horizon.  We should similarly expect an increase in the magnitude of the (negative) energy density near the black hole, as a buildup of negative-energy modes forms in the centrifugal box.  This physical reasoning leads us to conjecture that all components of the stress tensor should increase in magnitude as the angular momentum of the black hole is increased.

\subsection{Extremal Horizons}
\label{subsec:extremal}

Our family of solution black holes can be taken all the way to extremality.  This limit is particularly interesting, as the causal structure of an extremal black hole is qualitatively very different from its non-extremal relatives.  One might therefore imagine that the stress tensor of matter fields on extremal black hole spacetimes exhibits qualitatively different behavior as well.  In particular, the fact that the horizon of an extremal black hole is a Cauchy horizon would na\"ively lead one to believe that stress tensors should generically be singular on the future horizon of extremal black holes.  This question has been addressed by~\cite{Trivedi,Loranz,Balbinot:2004jx,Farese:2005nr}, who studied the regularity of the v.e.v. of the stress tensor outside an extremal dimensionally reduced two-dimensional Reissner-Nordstr\"om (RN) black hole.  In short, a static stress tensor (i.e. one sharing the same isometries as the background spacetime) exhibits a mild singularity on the future horizon, whereas the stress tensor of a massive scalar field propagating on an extremal two-dimensional RN geometry forming via gravitational collapse is regular thanks to the presence of decaying (but nonzero) flux terms.  As shown in~\cite{Anderson}, this subtle issue disappears in four dimensions, as the stress tensor of a scalar field in the zero-temperature vacuum state on the full four-dimensional RN spacetime is regular everywhere without the need for flux terms.

To our knowledge, the stress tensor of a field theory on a \textit{rotating} extremal black hole background has not yet been studied even in free field theory.  The four-dimensional RN results might lead us to expect that the stress tensor on extremal rotating black hole spacetimes should similarly be regular on the future horizon, though a key difference between the RN and rotating case is the inherent presence of matter in the RN spacetime.  In any case, the fundamental similarity between the causal structures of extremal charged and rotating black holes makes it quite plausible to expect that the CFT stress tensor dual to our rotating droplet will be regular on the future horizon even in the extremal limit.

\section{Constructing a Spinning Droplet}
\label{sec:numerics}

\subsection{The DeTurck Method}

The standard approach in AdS/CFT is to solve the vacuum Einstein field equations with negative cosmological constant\footnote{Here and below, we will use lower-case Latin letters to denote bulk indices, and lower-case Greek letters to denote boundary indices.}
\begin{equation}
\label{eq:EFE}
E_{ab} \equiv R_{ab} - \frac{1}{2} \, R g_{ab} + \Lambda g_{ab} = 0
\end{equation}
subject to certain boundary conditions.  The subleading behavior of the metric~$g_{ab}$ near the conformal boundary then contains information about the stress tensor of the dual CFT.  As is well known, Einstein's equations do not have a well defined character unless a gauge choice is made. Because the solutions we are searching for are stationary, one can hope to choose a clever gauge where the equations~(\ref{eq:EFE}), or a deformation thereof, are manifestly elliptic. This is exactly what the DeTurck trick does for us~\cite{DeTurck1983,Headrick:2009pv,Figueras:2011va}.  In short, one modifies the equations~\eqref{eq:EFE} by introducing a new vector~$\xi^a$ (called the DeTurck vector):
\begin{equation}
\label{eq:DeTurck}
E^H_{ab} \equiv E_{ab} - \grad_{(a} \xi_{b)} = 0, \ \ \ \xi^a = g^{bc} \left(\Gamma^a_{bc} - \bar{\Gamma}^a_{bc}\right),
\end{equation}
where~$\bar{\Gamma}^a_{bc}$ is the Levi-Civita connection of some reference metric~$\bar{g}$.  Equation~\eqref{eq:DeTurck} is called the Einstein-DeTurck or harmonic Einstein equation.  One can show that for stationary solutions with Killing horizons, the above choice of the DeTurck vector renders the Einstein-DeTurck equation elliptic.  Solving the Einstein-deTurck equations then reduces to solving a boundary-value problem. Note, however, that solutions to the Einstein-DeTurck equation are not solutions to the ordinary Einstein equations unless~$\xi^a = 0$. To get around this problem, one can show that the quantity~$\Phi \equiv \xi_a \xi^a$ must take its maximum value on the boundaries of the domain of integration; thus if the reference metric~$\bar{g}$ is chosen with the same boundary conditions as the metric~$g$,~$\Phi$ is zero on all boundaries of the integration domain, and must therefore be zero everywhere within as well~\cite{Figueras:2011va}. Then~$\xi^a$ is zero as well, and a solution to the Einstein-DeTurck equation is also a solution to the Einstein equations.  Indeed, the claim that the Einstein-DeTurck equation automatically takes care of gauge-fixing is justified by thinking of the condition~$\xi^a = 0$ as just a choice of gauge, which is a generalized harmonic gauge of the form $\Delta x^a = \bar{\Gamma}^a_{bc}g^{bc}$.

\subsection{Droplet Ansatz}

As mentioned at the end of Section~\ref{sec:intro}, the boundary metric we want to impose is the equal-angular-momentum~$d=5$ Myers-Perry solution~\cite{Myers1986304}:
\begin{subequations}
\label{eq:MP}
\begin{equation}
ds^2_\mathrm{MP} = -\frac{g(r)}{h(r)} \, dt^2 + \frac{dr^2}{g(r)} + r^2 \left[h(r)\left(d\psi + A_{(1)} + \Omega(r) dt\right)^2 + d\Sigma_2^2\right],
\end{equation}
\begin{align}
g(r) &\equiv \frac{(r^2 - r_0^2)(r^2 - \beta^2(r^2 + r_0^2))}{(1-\beta^2)r^4}, \\
h(r) &\equiv \frac{r^4 - \beta^2(r^4-r_0^4)}{(1-\beta^2)r^4}, \\
\Omega(r) &\equiv \frac{\beta r_0^3}{r^4 - \beta^2(r^4-r_0^4)},
\end{align}
\end{subequations}
where~$d\Sigma_2^2$ is the Fubini-Study metric on~$CP^1$,~$A_{(1)}$ is the K\"ahler potential of~$CP^1$, and~$\beta = r_0 \Omega_H$, with~$\Omega_H$ the angular velocity of the horizon.  Explicit forms for~$A_{(1)}$ and~$d\Sigma_2^2$ are given by the expressions
\begin{equation}
A_{(1)} = \frac{1}{2} \, \cos\theta \, d\phi, \ \ \ d\Sigma_2^2 = \frac{1}{4}\left(d\theta^2 + \sin^2\theta \, d\phi^2\right).
\end{equation}

For~$\beta = 0$, the metric~\eqref{eq:MP} reduces to~$d = 5$ Schwarzschild~\eqref{eq:Schwd} with the~$S^3$ written as a Hopf fibration of~$S^1$ over~$CP^1$; setting~$\beta^2 = \beta^2_\mathrm{ext} \equiv 1/2$ yields the extremal solution.  The metric is cohomogeneity one owing to its~$U(2)$ symmetry, which implies that our bulk solution will be cohomogeneity two.

Let us compactify the range of the radial coordinate~$r$ to yield a metric more amenable to our numerical approach.  We also rescale~$t \rightarrow r_0 \, t$ and exploit the coordinate freedom $\psi \rightarrow \psi + \lambda t$ to shift~$\Omega$ so that~$\Omega(r_0) = 0$,~$\Omega(\infty) = -\Omega_H$ (this will simplify the boundary conditions at the horizon later on).  Defining a new radial coordinate~$y = 1-(r_0/r)^2$, we find
\begin{subequations}
\label{eq:MPcompact}
\begin{multline}
ds^2_\mathrm{MP} = r_0^2\left\{-y \, \frac{h_1(y)}{h_2(y)} \, dt^2 + \frac{(1-\beta^2) dy^2}{4 y \, h_1(y)(1-y)^3} \right. \\ \left. + \frac{1}{1-y} \left[\frac{h_2(y)}{1-\beta^2}\left(d\psi + A_{(1)} - \beta(1-\beta^2) \, \frac{y (2-y)}{h_2(y)} \, dt\right)^2 + d\Sigma_2^2\right]\right\},
\end{multline}
\begin{align}
h_1(y) &\equiv 1-\beta^2(2-y), \\
h_2(y) &\equiv 1-\beta^2 y(2-y).
\end{align}
\end{subequations}
Our ansatz will need to have~\eqref{eq:MPcompact} as its conformal boundary.

Next, recall that we are considering black droplet solutions with~$T_\infty = 0$; this implies that far into the bulk, the solution should approach the near-horizon geometry of pure~AdS$_6$ in Poincar\'e coordinates.  Poincar\'e AdS$_6$ in standard coordinates can be written as
\begin{equation}
ds^2_\mathrm{Poin} = \frac{\ell^2}{z^2}\left[-dt^2 + dz^2 + dr^2 + r^2 d\Omega_3^2\right],
\end{equation}
where~$\ell$ is the AdS$_6$ length scale, related to the cosmological constant by~$\ell^2 = -10/\Lambda_6$.  Changing variables to~$(x,y)$ defined by
\begin{equation}
z = \frac{1-x^2}{\sqrt{1-y}}, \ \ \ r = x\sqrt{\frac{2-x^2}{1-y}},
\end{equation}
the Poincar\'e metric becomes
\begin{equation}
\label{eq:Poin}
ds^2_\mathrm{Poin} = \frac{\ell^2}{(1-x^2)^2}\left[-(1-y) \, dt^2 + \frac{dy^2}{4(1-y)^2} + \frac{4 \, dx^2}{2-x^2} + x^2(2-x^2) d\Omega_3^2\right].
\end{equation}
The near-horizon region corresponds to taking~$y$ close to~1.

We are now ready to write down an ansatz.  Consider
\begin{multline}
\label{eq:ansatz}
ds^2 = \frac{\ell^2(1-y)}{(1-x^2)^2} \left\{-y \, T \, \frac{h_1(y)}{h_2(y)} \, dt^2 + \frac{A \, dy^2}{4 y h_1(y)(1-y)^3} \right. \\ \left.  + \frac{4B}{(1-y)(2-x^2)}\left(dx + F \, \frac{x(1-x^2)}{2(1-y)} \, dy \right)^2 \right. \\ \left. + \frac{x^2(2-x^2)}{1-y} \left[\tilde{h}_2(x,y) C \left(d\psi + A_{(1)} - G \, \frac{y (2-y)}{h_2(y)} dt\right)^2 + S \, d\Sigma_2^2\right]\right\},
\end{multline}
where
\begin{equation}
\tilde{h}_2(x,y) \equiv 1 + x^2\left(\frac{h_2(y)}{1-\beta^2} - 1\right),
\end{equation}
$T$,~$A$,~$B$,~$C$,~$S$,~$F$, and~$G$ are all functions of~$x$ and~$y$, and the coordinate range is the rectangle~$(x,y) \in (0,1) \times (0,1)$.  For the non-extremal case~$\beta \neq 1/\sqrt{2}$, our boundary conditions are as follows:
\begin{itemize}
\item At the conformal boundary~$x \rightarrow 1$, we take~$T = A/(1-\beta^2) = C = S = r_0^2/\ell^2$,~$B = 1$,~$G = \beta(1-\beta^2)$, and~$F = 0$.  Then the metric approaches
	\begin{equation}
	ds^2 \rightarrow \frac{\ell^2}{(1-x)^2} \left(dx^2 + \frac{1-y}{4\ell^2} \, ds^2_\mathrm{MP}\right)
	\end{equation}
	as desired.
\item Near the extremal Poincar\'e horizon~$y \rightarrow 1$, we take~$T = A/(1-\beta^2) = B = C = S = 1$,~$G = \mathrm{const.} = \beta(1-\beta^2)$, and~$F = 0$ (note that the constant value of~$G$ is fixed by requiring consistency with the boundary condition at~$x = 1$).  Changing to a new coordinate~$\psi \rightarrow \psi + \beta t$ removes the~$dt$ cross-terms, and we recover the~$y \rightarrow 1$ limit of~\eqref{eq:Poin}.
\item At the horizon~$y \rightarrow 0$, we require regularity of the metric, which imposes a relationship between~$T$ and~$A$.  In particular, we must have~$T/A = (r_0 \kappa/(1-2\beta^2))^2$, where~$\kappa = (1-2\beta^2)/(r_0\sqrt{1-\beta^2})$ is the surface gravity of the boundary black hole.  Thus we find~$T/A = 1/(1-\beta^2)$.  In addition, expanding the equations of motion and the condition~$\xi^a = 0$ order-by-order near~$y = 0$ gives Robin conditions on~$\partial_y M_i|_{y = 0}$ for~$M_i = \{T,A,B,C,S,F,G\}$.  
\item At the fixed point of the~$U(2)$ isometry~$x = 0$, we require regularity as well.  Again, this implies that all metric functions must be smooth functions of~$x^2$, giving the Neumann conditions~$\partial_x M_i|_{x = 0} = 0$.  In addition, note that while the forms~$x^2 (d\psi + A_{(1)})$ and~$x dx$ are regular at~$x = 0$, in general the metric components
	\begin{equation}
	2 \, B \, dx^2 + 2 \, x^2 \left[\tilde{h}_2(x,y) C \left(d\psi + A_{(1)}\right)^2 + S \, d\Sigma_2^2\right]
	\end{equation}
	are not.  In order to make the above expression regular, we must impose that~$B = C = S$ at~$x = 0$, so that
	\begin{equation}
	2 \, B \, dx^2 + 2 \, x^2 \left[\tilde{h}_2(x,y) C \left(d\psi + A_{(1)}\right)^2 + S \, d\Sigma_2^2\right] \rightarrow 2S (dx^2 + x^2 d\Omega_3^2),
	\end{equation}
	which is manifestly regular there.
\end{itemize}
Having understood the boundary conditions on each side of the computational rectangle, we choose the reference metric.  Noting from~\eqref{eq:MPcompact} that the boundary black hole size~$r_0$ enters only as a conformal factor, we can choose without loss of generality to set~$r_0 = \ell$.  Then the boundary conditions become consistent, and we take the reference metric to be the ansatz~\eqref{eq:ansatz} with~$T = A/(1-\beta^2) = B = C = S = 1$,~$G = \beta(1-\beta^2)$, and~$F = 0$.

The extremal case ($\beta = \beta_\mathrm{ext} \equiv 1/\sqrt{2}$) takes a little more care.  Though our ansatz~\eqref{eq:ansatz} still applies, one finds that as extremality is approached the metric function~$F$ attains a very large value along~$y = 0$ near~$x = 0$, leading to large numerical errors.  To alleviate the problem, we rewrite~\eqref{eq:ansatz} in terms of a new metric function~$\widetilde{F} = yF$; we then find that the numerics are much more well-behaved when written in terms of the new~$\widetilde{F}$ rather than the old~$F$.  Then the boundary conditions at the conformal boundary, fixed point of the~$U(2)$ isometry, and extremal Poincar\'e horizon remain the same as those discussed above, while at the extremal droplet horizon we require~$\tilde{F} = 0$.  In addition, converting to Eddington-Finkelstein coordinates and requiring regularity at the extremal droplet horizon imposes the usual regularity condition~$A(x,0) = (1-\beta_\mathrm{ext}^2)T(x,0) = T(x,0)/2$, as well as the condition~$G(x,0) = \mathrm{const.} = \beta_\mathrm{ext}(1-\beta_\mathrm{ext}^2) = 1/\sqrt{8}$.  Boundary conditions on the other metric functions are obtained by expanding the equations of motion order-by-order near~$y = 0$ and requiring that they vanish to leading order.

\subsection{Numerics}

\begin{figure}[t]
\centering
\includegraphics[width=0.4\textwidth]{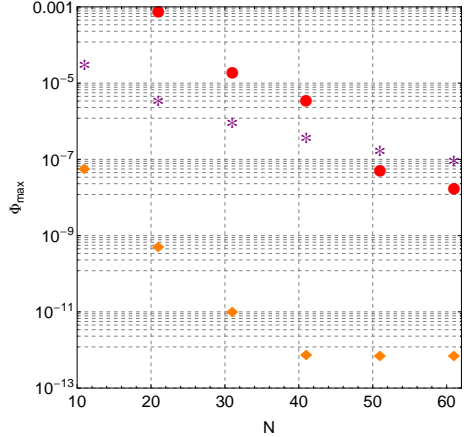}
\caption{The maximum value of the square of the DeTurck vector~$\Phi$ as a function of~$N$ for~$\beta = 0$ (circles),~$0.6$ (diamonds), and~$\beta_\mathrm{ext} \approx 0.707$ (asterisks).}
\label{fig:Phi}
\end{figure}

We approximate the Einstein-DeTurck equation~\eqref{eq:DeTurck} with the ansatz~\eqref{eq:ansatz} using pseudospectral collocation methods on a Chebyshev grid.  The resulting nonlinear algebraic equations are solved using a damped Newton-Raphson method.  We monitor the damping with~$|E^H_{ab}|$; that is, we ensure that each iteration of the Newton's method decrease the magnitude of~$E^H_{ab}$.  Due to the maximum principle that~$\Phi \equiv \xi_a \xi^a$ obeys, we can monitor the error in our solutions by monitoring the maximum value of~$\Phi$.  In Figure~\ref{fig:Phi}, we plot the maximum value of~$\Phi$ as a function of the number~$N$ of grid points for three different values of the rotation parameter~$\beta$, showing the expected exponential decrease.  Note that we will find later that the accuracy of the extracted of the stress tensor depends strongly on the number of grid points used along the~$x$-direction, so for all results shown in this paper, we use an~$N_x \times N_y = 81 \times 41$ grid.

\subsection{Extraction of the Stress Tensor}

In order to extract the boundary stress tensor from our solutions, we follow the approach used in~\cite{Santos:2012he}.  In short, we first expand the metric functions~$M_i$ as a power series in~$x$ off of the boundary at~$x = 1$:
\begin{equation}
\label{eq:expansion}
M_i(x,y) = \sum_{n = 0} M_i^{(n)}(y) (1-x)^n + \ln(1-x) \sum_{n=0} \widetilde{M}_i^{(n)}(y)(1-x)^n,
\end{equation}
where the logarithmic terms only appear for~$\beta \neq 0$ and at no lower order than~$n = 5$.  Inserting these expansions into~\eqref{eq:DeTurck} and in addition imposing~$\xi^a = 0$ allows us to solve for the coefficients of the power series order-by-order in~$(1-x)$.  The expansion is unique up to~$n = 4$ for the functions~$\{T,A,C,S,F,G\}$ and up to~$n = 5$ for the function~$B$.  We then change to Fefferman-Graham coordinates~$(\tilde{z},\tilde{y})$ using an expansion of the form
\begin{subequations}
\label{eq:FGexp}
\begin{align}
1- x^2 &= \sqrt{1-\tilde{y}} \, \tilde{z} + \sum_{n=2} X^{(n)}(\tilde{y}) \tilde{z}^n, \\
y &= \tilde{y} + \sum_{n=1} Y^{(n)}(\tilde{y}) \tilde{z}^n,
\end{align}
\end{subequations}
where the~$\mathcal{O}(\tilde{z})$ term in the expansion for~$x$ fixes the conformal frame, and we have neglected potential logarithms as they do not appear up to the order we need.  Requiring that $g_{\tilde{z}\tilde{z}} = \ell^2/\tilde{z}^2$ and~$g_{\tilde{z} \tilde{y}} = 0$ order-by-order in~$\tilde{z}$ yields a set of algebraic equations that can be solved for the coefficients~$X^{(n)}(\tilde{y})$,~$Y^{(n)}(\tilde{y})$.  Then we find that the metric takes the form
\begin{equation}
\label{eq:FG}
ds^2 = \frac{\ell^2 }{\tilde{z}^2}\left[d\tilde{z}^2 + \ell^{-2} ds^2_\mathrm{MP} + \tilde{z}^5 h_{\mu\nu} \, dx^\mu \, dx^\nu + \mathcal{O}(\tilde{z})^6\right],
\end{equation}
where~$h_{\mu\nu}$ is only a function of the boundary coordinates~$x^\mu$.  According to the prescription of~\cite{deHaro:2000xn}, the renormalized v.e.v.~of the boundary stress tensor is then
\begin{equation}
\langle T_{\mu\nu} \rangle = \frac{5\ell^4}{16\pi G_N^{(6)}} \, h_{\mu\nu}.
\end{equation}

The general expressions for the coefficients~$M_i^{(n)}$,~$\widetilde{M}_i^{(n)}$,~$X^{(n)}$,~$Y^{(n)}$ and for the stress tensor are too cumbersome to reproduce in this paper, but they become tractable in the nonrotating case~$\beta = 0$.  In Appendix~\ref{ap:expansion}, we give the~$\beta = 0$ expressions for~$M_i^{(n)}$,~$X^{(n)}$, and~$Y^{(n)}$ up to the order we need in order to extract the stress tensor and check its tracelessness and transversality. As expected from our arguments in Section~\ref{subsec:SchwTmunu}, the stress tensor is given precisely by the expression~\eqref{eq:T3}, with\footnote{Note that we drop the tilde on~$\tilde{y}$, since~$y$ and~$\tilde{y}$ agree on the boundary.}
\begin{subequations}
\label{eqs:ThetaG}
\begin{align}
\Theta(y) &= -\frac{5\ell^4}{16\pi G_N^{(6)}} \, \frac{(1-y)^{5/2}\left[14(1-2y) a_5(y) + (1-y)(176-1712y+2368y^2 + 14ya_5'(y)) \right]}{672(1-2y)}, \\
G(y) &= \frac{5\ell^4}{16\pi G_N^{(6)}}\, \frac{y(1-y)\left(88-384y+296y^2+7a_5(y)\right)}{224},
\end{align}
\end{subequations}
where in an obvious abuse of notation we have written the functions~$\Theta$ and~$G$ in terms of the compactified radial coordinate~$y$ rather than the original coordinate~$r$.  We will postpone a discussion of this result to Section~\ref{sec:disc}.

Now let us return to the general~$\beta \neq 0$ case and discuss how to extract the stress tensor from our numerical data.  The boundary stress tensor will take the form
\begin{equation}
\label{eq:Tmunu}
\langle T_{\mu\nu}\rangle dx^\mu d^\nu = T_{tt} \, dt^2 + T_{rr} \, dr^2 + T_{\psi\psi} \left(d\psi + A_{(1)}\right)^2 + 2T_{t\psi} \left(d\psi + A_{(1)}\right) dt + T_{\Sigma\Sigma} d\Sigma_2^2.
\end{equation}
There are thus five functions to solve for, which can be expressed in terms of five of the functions~$M_i^{(5)}$ which we will label as $m^{(5)}_i \equiv \{t_5, a_5, c_5, g_5, s_5\}$.  In principle, one could extract these functions by taking five~$x$-derivatives of~\eqref{eq:expansion} at~$x = 1$.  However, in practice taking such high derivatives numerically gives very poorly behaved results.  Instead, we follow a slightly different approach: since we know the coefficients~$M_i^{(n)}$ for~$n \leq 4$ analytically, we can write
\begin{equation}
\label{eq:extractcoeffs}
\frac{M_i(x,y) - \sum_{n=0}^4 M_i^{(n)}(y) (1-x)^n}{(1-x)^4} = m^{(5)}_i(y)(1-x) + \mathcal{O}(1-x)^2.
\end{equation}
With the numerically computed values for~$M_i(x,y)$, the left-hand side of~\eqref{eq:extractcoeffs} is known numerically, so we perform a fit for the right-hand side to obtain the coefficients~$m^{(5)}_i$.

\begin{figure}[t]
\centering
\includegraphics[width=0.4\textwidth]{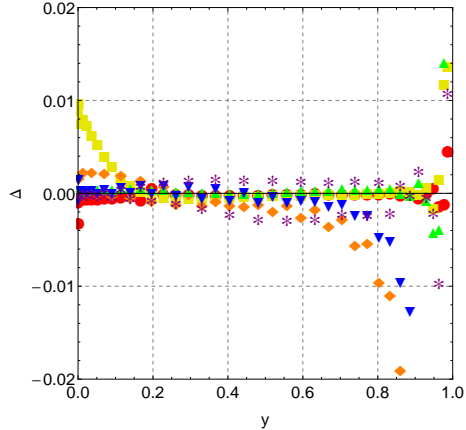}
\caption{The relative error~$\Delta$ introduced in extracting the coefficients~$m^{(5)}_i$ for different values of~$\beta$.  Circles, squares, diamonds, triangles, inverted triangles, and asterisks correspond to~$\beta = 0$,~$0.5$,~$0.6$,~$0.65$,~$0.69$, and~$\beta_\mathrm{ext}$ respectively.  Note that the error remains~$\lesssim 0.4$\% for almost all~$\beta$ and~$y$.}
\label{fig:Delta}
\end{figure}

In order to quantify the accuracy of our extracted stress tensor, we proceed as follows.  Although the equations of motion don't give unique expressions for the coefficients~$m^{(5)}_i$, they do provide two algebraic relations between these coefficients and~$a_5'$\footnote{These relations enforce tracelessness and transversality of the stress tensor, \emph{i.e.} conservation.}.  Using one of these relations, we can express~$t_5$ in terms of~$a_5$,~$c_5$, and~$g_5$ in an algebraic expression of the form
\begin{equation}
t_5(y) = \mathcal{T}_5^\mathrm{exact}\left(y,a_5(y),c_5(y),s_5(y)\right).
\end{equation}
Using the fitted values for the~$m^5_i$, we can therefore use the quantity
\begin{equation}
\Delta(y) \equiv 1 - \frac{t_5(y)}{\mathcal{T}_5^\mathrm{exact}\left(y,a_5(y),c_5(y),s_5(y)\right)},
\end{equation}
as a measure of the error introduced in extracting the~$m^{(5)}_i$ (physically, one can think of~$\Delta$ as a measure of how much the numerically extracted stress tensor fails to be traceless).  In Figure~\ref{fig:Delta}, we plot~$\Delta$ for the full range of~$\beta$. We note that the relative error in our extraction is~$\lesssim 0.4$\% for almost all~$\beta$ and~$y$; only for~$y$ near~1 does~$\Delta$ become appreciable.  We emphasize that this error is solely due to the difficulty in extracting the~$m_i^{(5)}$ from the numerical data, and is not a measure of the accuracy of our numerical solutions themselves.

\begin{figure}[h!]
\centering
\includegraphics[width=0.8\textwidth]{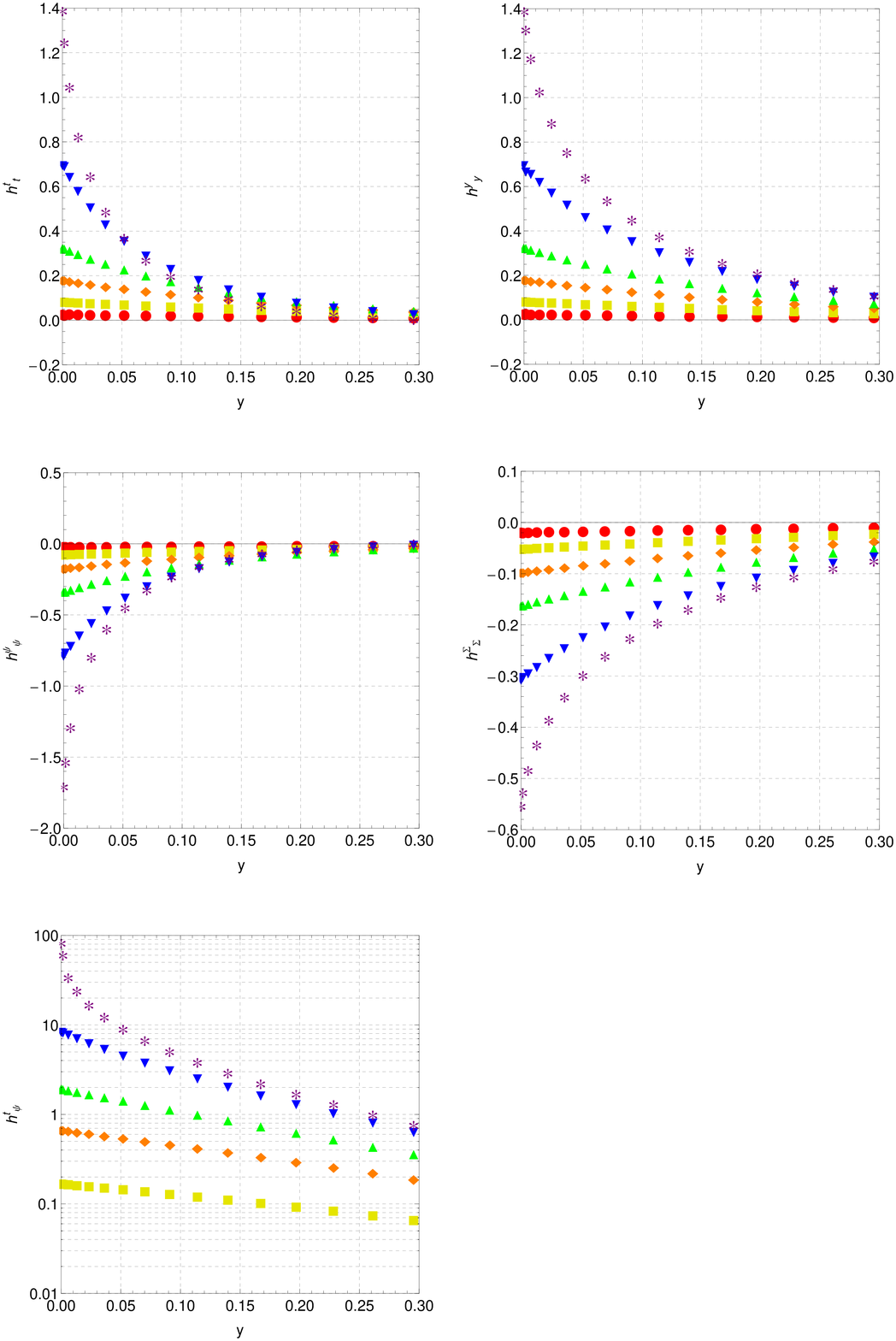}
\caption{The components~${h^\mu}_\nu = 16\pi G_N^{(6)} \left\langle {T^\mu}_\nu \right\rangle / 5\ell^2$ for the same values of~$\beta$ as Figure~\ref{fig:Delta}.  Note that~${h^t}_\psi = 0$ for~$\beta = 0$.  Note that the range of the coordinate~$y$ shown corresponds to the range~$r_0 \leq r \lesssim 1.2 \, r_0$ in the original radial coordinate.}
\label{fig:stresstensor}
\end{figure}

\begin{figure}[t]
\centering
\includegraphics[width=0.4\textwidth]{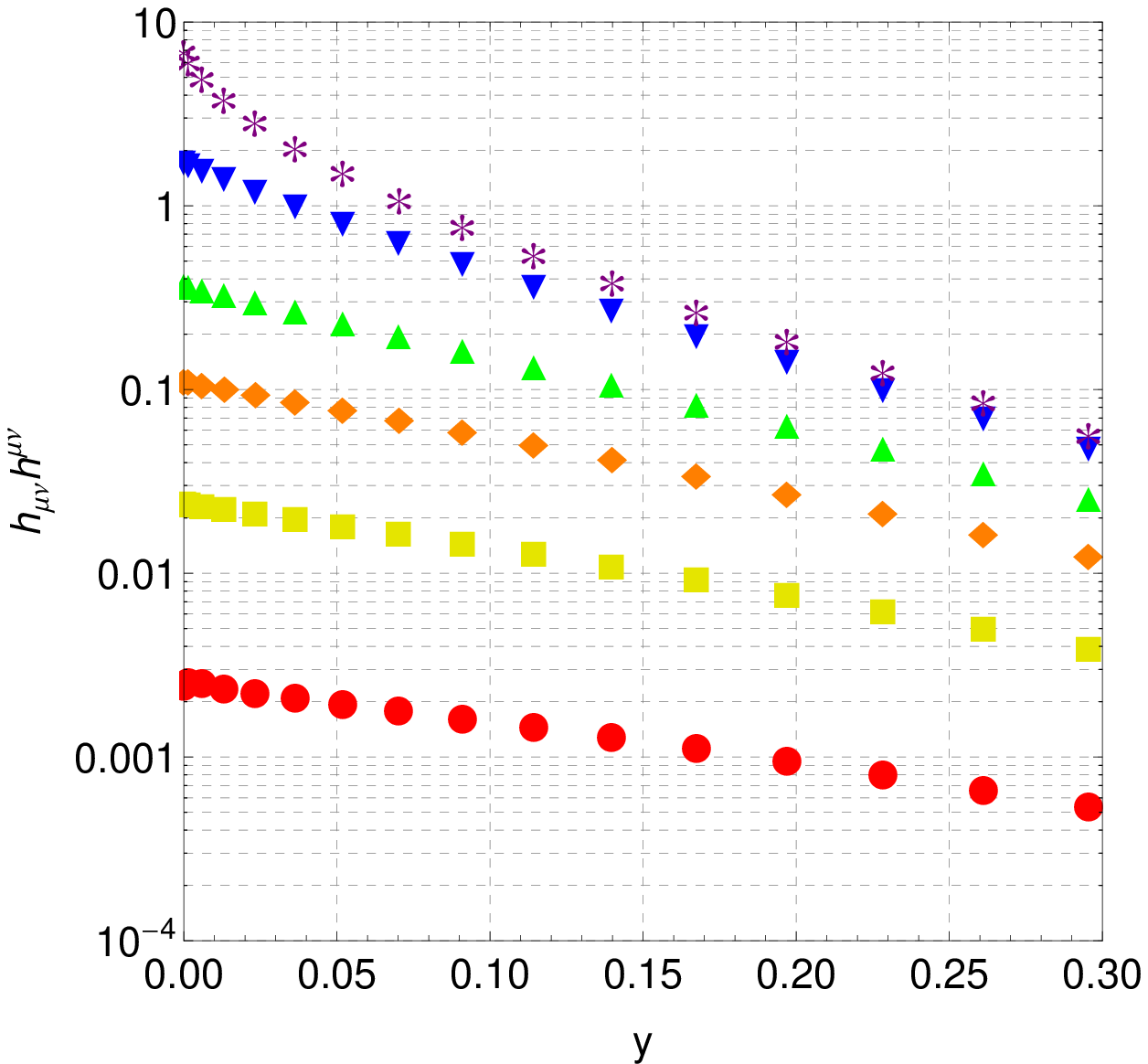}
\caption{The scalar invariant~$h_{\mu\nu} h^{\mu\nu}$ for the same values of~$\beta$ as Figure~\ref{fig:Delta}.}
\label{fig:Tsquared}
\end{figure}

In Figure~\ref{fig:stresstensor}, we plot the components of the stress tensor for various values of~$\beta$.  Recall that to obtain the metric~\eqref{eq:MPcompact}, we shifted the angular coordinate~$\psi$ to make~$g_{\psi t} = 0$ at the horizon.  As a result, the near-horizon behavior shown in Figure~\ref{fig:stresstensor} physically represents the stress tensor as measured by an observer co-rotating with the horizon.  In addition, in Figure~\ref{fig:Tsquared} we plot the scalar invariant~$\left\langle T_{\mu\nu}\right\rangle \left\langle T^{\mu\nu} \right\rangle$.

\subsubsection{Regularity at Horizon}

From the discussion in Section~\ref{subsec:SchwTmunu} and the regularity of the functions~\eqref{eqs:ThetaG}, it is clear that in the nonrotating case, the stress tensor of the CFT is regular on the future horizon (and in fact, it is regular on the past horizon as well since~$K = Q = 0$).  One can similarly examine the behavior of the stress tensor in the rotating case by passing to local ingoing Eddington-Finkelstein coordinates.  In the non-extremal case, we change coordinates to
\begin{equation}
dt = dv - \frac{\sqrt{1-\beta^2}}{1-2\beta^2} \, \frac{dy}{2y};
\end{equation}
then from~\eqref{eq:Tmunu}, one finds that the stress tensor is regular on the future horizon only if all the components~$\left\langle{T^\mu}_\nu\right\rangle$ are finite and satisfy~$\left\langle{T^t}_t\right\rangle = \left\langle{T^y}_y\right\rangle$ on the horizon.  We can immediately see from Figure~\ref{fig:stresstensor} that these conditions are satisfied, so the stress tensor is regular on the non-extremal horizon.

Again, the extremal case requires more care.  In this case, local ingoing Eddington-Finkelstein coordinates are given by
\begin{subequations}
\begin{align}
dt &= dv - \frac{1}{\sqrt{2}}\left(\frac{1}{y^2} + \frac{1}{y}\right) \, dy, \\
d\psi &= d\psi' - \frac{dy}{2 y}.
\end{align}
\end{subequations}
Then one again finds that the stress tensor is regular on the future horizon if it obeys the same conditions as in the non-extremal case, and in addition obeys
\begin{equation}
\sqrt{2} \left\langle {T^t}_\psi \right\rangle + 2 \partial_y \left(\left\langle {T^t}_t \right\rangle - \left\langle {T^y}_y \right\rangle\right) = 0
\end{equation}
on the horizon.  It is easy to check that this condition is satisfied to within the expected~$\sim 1 \%$ accuracy.

\section{Results and Discussion}
\label{sec:disc}

Our numerical results exhibit the behavior expected based on our quantitative and heuristic reasoning in Section~\ref{sec:analytic}.  Let us begin with the nonrotating case~$\beta = 0$.  The expansion of the equations of motion off of the boundary gave us a stress tensor of the form~\eqref{eq:T3} with the functions~$\Theta$ and~$G$ given in~\eqref{eqs:ThetaG}, which is precisely the form expected from the discussion in Section~\ref{subsec:SchwTmunu}.  In particular, we find that the stress tensor is traceless, contains no flux terms, and is regular at the future horizon.  Once the numerically extracted functions~$m_i^5(y)$ are inserted, we find that the energy density is negative everywhere.  The magnitude of the components of the stress tensor increases as the rotation parameter is increased, but otherwise the qualitative behavior of the stress tensor doesn't change when~$\beta \neq 0$ except for the introduction of an angular flux term~$T_{t\psi}$.  

As shown in Figure~\ref{fig:Delta}, the error in our extracted stress tensor becomes relatively large near~$y = 1$, making it difficult to reliably extract the falloff of the stress tensor components at large~$r$.  In fact, these components appear to exhibit a falloff closer to~$(1-y)^3 = (r_0/r)^6$ than to the~$(1-y)^{7/2} = (r_0/r)^7$ expected from the braneworld arguments of~\cite{Figueras:2011va}.  However, by performing an expansion of the equations of motion about~$y = 1$, one can show analytically that the stress tensor components must indeed decay like~$(r_0/r)^7$; presumably, the use of higher precision would allow us to extract this behavior numerically.  In any case, our results are much more well-behaved in the near-horizon region, where most of the interesting physics lies.

The following physical pictures emerges.  As expected, a strongly coupled large-$N$ CFT in a jammed phase forms a halo of negative-energy plasma around a black hole.  The stress tensor of the plasma falls off rapidly far from the black hole, indicating that the plasma is well-localized around the black hole.  The total energy of the plasma is negative, indicating its highly non-classical nature.  When the black hole is made to rotate, the plasma is dragged along with the black hole.  The plasma is trapped by the centrifugal barrier created by this rotation, causing the energy density and pressures to increase in magnitude.

The jammed phase is particularly interesting from the point of view of the CFT because it has no analog in a free field theory: a similar static plasma localized around the black hole could not exist in a noninteracting theory, as it would quickly fall into the black hole. The jammed phase is thus an effect of the strongly coupled nature of the CFT.

One might therefore wonder how our droplet phase compares to the corresponding black funnel phases (which \textit{do} admit analogs in free field theory).  Black funnels with~$T_\infty = 0$ must have a horizon that asymptotes to the extremal Poincar\'e AdS deep into the bulk; if we require that the bulk horizon also join smoothly to the (non-extremal) boundary horizon, the bulk horizon must have both a non-constant surface gravity and a non-constant angular velocity, and will therefore not be a Killing horizon.  We say that such funnel solutions will ``flow'' in the sense that there exists some notion of a horizon velocity which moves either from the boundary black hole to infinity or vice versa.  For non-extremal horizons, these flowing funnels were constructed in~\cite{Fischetti:2012vt}, but construction of solutions with asymptotically extremal horizons (as is necessary to have the boundary CFT be at zero temperature at infinity) is more difficult.

In fact, it would seem that such funnel solutions might not even be relevant: according to the discussion in Section~\ref{sec:intro}, the conjectured jamming transition in the CFT should be parametrized by the parameter~$RT_\infty$, with small~$RT_\infty$ favoring the jammed phase.  If~$T_\infty = 0$, we might expect based on this argument that the jammed phase of the CFT will dominate the thermodynamic ensemble for any size of the boundary black hole.

However, recall that one basis for this conjectured phase transition was the potential for Gregory-Laflamme-type instabilies of the bulk horizon~\cite{Gregory:2000gf}: if one starts with a black funnel and lowers the temperature~$T_\infty$, the asymptotically planar bulk horizon will sink deeper into the bulk (cf. Figure~\ref{fig:dropfun}).  The na\"ive expectation is that this will narrow the neck of the black funnel and leave it unstable to Gregory-Laflamme instabilities, causing it to collapse into a black droplet.  Taking~$T_\infty \rightarrow 0$, this argument would lead one to expect that the droplet solution would be the only thermodynamically preferred one.

\begin{figure}[t]
\centering
\begin{pspicture}(6,4)

\pscustom{
\gsave
	\psbezier(0,0)(1,1)(2,2)(2,4)
	\psline(4,4)
	\psbezier(4,4)(4,2)(5,1)(6,0)
	\psline(0,0)
	\fill[fillstyle=solid,fillcolor=lightgray]
\grestore
}

\psline(0,4)(6,4)
\psbezier[showpoints=false](2,4)(2,2)(1,1)(0,0)
\psbezier[showpoints=false](4,4)(4,2)(5,1)(6,0)

\psdots(2,4)(4,4)

\end{pspicture}
\caption{A sketch of a black funnel solution with~$T_\infty = 0$.}
\label{fig:T0funnel}
\end{figure}
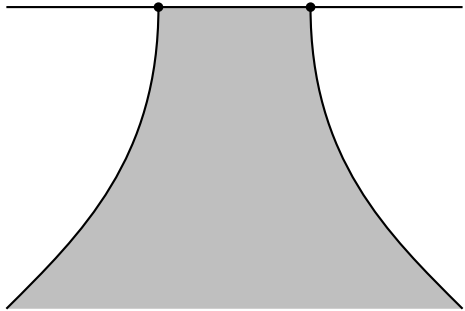

This potential instability was studied to some extent in~\cite{Santos:2012he}, which considered funnels dual to a CFT on an asymptoticaly flat black hole spacetime. The size of the boundary black hole was decreased (while keeping the temperature~$T_\infty$ fixed) to explore whether such Greogory-Laflamme instabilities would occur.  Surprisingly, as the size of the boundary black hole decreased, the neck of the funnel remained wide enough to prevent the formation of such instabilities.  Though far from exhaustive, this result implies that it is possible for there to be stable funnels with~$T_\infty = 0$ which could compete with the droplet solutions.  A sketch of what such a solution might look like is shown in Figure~\ref{fig:T0funnel} (in fact, in~\cite{Fischetti:2012ps} such funnels were explicitly constructed in~$d = 2$, though the question of their stability is less interesting in that case, since the droplet solutions cannot exist in three bulk dimensions).  Clearly, the construction of such solutions would be of much interest.  We leave this exploration to future work.

Some words on the extremal limit are also in order. Though this limit is discontinuous from the point of view of the global causal structure of the background geometry, the line element~\eqref{eq:ansatz} and consequently the stress tensor change smoothly as~$\beta$ approaches~$\beta_\mathrm{ext}$.  In particular, the stress tensor remains regular on the future horizon, consistent with the result of~\cite{Anderson} for the four-dimensional RN spacetime.  A natural question is then whether there is some other mechanism that prevents the spacetime from continuing past the Cauchy horizon, analogous to the mass-inflation singularity of black holes with inner horizons~\cite{Poisson1990}; perhaps the inclusion of perturbative corrections at higher order in~$1/N$ would resolve the issue. 

Indeed, our results may also have implications for the mass-inflation singularity.  Ref.~\cite{Steif:1993zv} computed the stress tensor on a rotating BTZ black hole background and found that it diverged at the inner horizon.  This result should be expected in any black hole spacetime with an inner (Cauchy) horizon to preserve the causal structure of the spacetime.  Since the spacetime between the inner and outer horizons is dynamical, the numerical approach used in this paper would not be applicable to studying the event horizon in that region.  However, obtaining the stress tensor of a field theory on a black hole background all the way to the inner horizon would certainly give some insight into the behavior of the mass inflation singularity there.

\subsection*{Acknowledgements}
We thank Netta Engelhardt, Aron Wall, Benson Way, and most especially Don Marolf for stimulating discussions about rotating droplets and their concomitant phase diagram. This work was supported by the National Science Foundation under Grant No.s PHY08-55415 and PHY12-05500.

\appendix

\section{Near-Boundary Expansion for the Nonrotating Droplet}
\label{ap:expansion}

In this appendix, we list the expansions of the metric functions near the conformal boundary~$x = 1$ for the nonrotating~($\beta = 0$) droplet, as well as the form of the Fefferman-Graham expansion~\eqref{eq:FG}.  For the metric functions, we have
\begin{subequations}
\begin{multline}
T(x,y) = 1 + (1-y)\left(-6(1-x)^2 + 6(1-x)^3 + \frac{67-408y^2}{14} \, (1-x)^4\right)  \\ + t_5(y)(1-x)^5 + \mathcal{O}(1-x)^6,
\end{multline}
\begin{multline}
A(x,y) = 1 + (1-y)\left(-6(1-x)^2 + 6(1-x)^3 + \frac{67-296y^2}{14} \, (1-x)^4\right)  \\ + a_5(y)(1-x)^5 + \mathcal{O}(1-x)^6,
\end{multline}
\begin{multline}
S(x,y)  = C(x,y) = 1 + (1-y)\left(2(1-x)^2 - 2(1-x)^3 + \frac{5(19+8y^2)}{14} \, (1-x)^4\right) \\ + s_5(y)(1-x)^5 + \mathcal{O}(1-x)^6,
\end{multline}
\begin{align}
B(x,y) &= 1 + \frac{4}{7} \, (1-y)(5+y)\left((1-x)^4 - 2 (1-x)^5\right) + \mathcal{O}(1-x)^6, \\
F(x,y) &= -(1-y)(1-x)^2 \left[1 + (1-x) + \frac{211-192y}{28} \, (1-x)^2 + \mathcal{O}(1-x)^3\right], \\
G(x,y) &= 0.
\end{align}
\end{subequations}
The coefficients~$t_5(y)$,~$a_5(y)$, and~$s_5(y)$ cannot be determined uniquely, but must satisfy the relationships
\begin{subequations}
\begin{align}
t_5(y) &= \frac{7(1-2y)a_5(y) - (1-y)(656y - 960y^2 - 14ya_5'(y))}{7(1-2y)}, \\
s_5(y) &= -\frac{14(1-2y)a_5(y) + (1-y)(8(55 - 265 y + 266y^2) + 14ya_5'(y))}{21(1-2y)}.
\end{align}
\end{subequations}
For the transformation to Fefferman-Graham coordinates, we have
\begin{subequations}
\begin{equation}
1 - x^2 = \tilde{z} \sqrt{1-\tilde{y}} \left[1 - \frac{1-\tilde{y}^2}{4} \,\tilde{z}^2 + \frac{3(1-\tilde{y})^2 (3+6\tilde{y} + 19\tilde{y}^2)}{224} \, \tilde{z}^4 + \mathcal{O}(\tilde{z})^6 \right],
\end{equation}
\begin{multline}
y = \tilde{y}\left[1 + (1-\tilde{y})^2 \, \tilde{z}^2 + \frac{(1-\tilde{y})^3 (1-3\tilde{y})}{2} \, \tilde{z}^4 \right. \\ \left. - \frac{(1-\tilde{y})^4 (4 + 19\tilde{y} - 44\tilde{y}^2)}{21} \, \tilde{z}^6 + \mathcal{O}(\tilde{z})^7\right].
\end{multline}
\end{subequations}


\bibliographystyle{JHEP}
\bibliography{Droplet3}

\end{document}